\documentclass[journal=jctcce]{achemso}

\usepackage{mathpazo}
\usepackage{amssymb}
\usepackage[fleqn]{amsmath}
\usepackage{graphicx}
\usepackage{epstopdf}
\usepackage[titletoc,title]{appendix}
\usepackage{gensymb}
\usepackage{bm}

\usepackage{xcolor}

\author{J\'er\'emy R. Rouxel}
\email{jrouxel@uci.edu}
\affiliation[UCI]{Department of Chemistry and Department of Physics and Astronomy, University of California, Irvine, CA 92697}

\author{Ahmadreza Rajabi}
\affiliation[UCI]{Department of Chemistry and Department of Physics and Astronomy, University of California, Irvine, CA 92697}

\author{Shaul Mukamel}
\email{smukamel@uci.edu}
\affiliation[UCI]{Department of Chemistry and Department of Physics and Astronomy, University of California, Irvine, CA 92697}

\date{\today}

\title{Chiral Four-Wave-Mixing signals with circularly-polarized X-ray pulses}

\begin{document}

\maketitle

\begin{abstract}
Chiral four-wave-mixing signals are calculated using the irreducible tensor formalism.
Different polarization and crossing angle configurations allow to single out the magnetic dipole and the electric quadrupole interactions.
Other configurations can reveal that the chiral interaction occurs at a given step within the nonlinear interaction pathways contributing to the signal.
Applications are made to the study of valence excitations of S-ibuprofen by chiral Stimulated X-ray Raman signals at the Carbon K-edge and by chiral visible 2D Electronic Spectroscopy.
\end{abstract}

\section{Introduction}

Chirality is the notable property of molecules lacking mirror symmetry.
This simple geometrical constraint has profound implications on fundamental science\cite{quack2002important}, biological activity\cite{flugel2011chirality} and drug synthesis\cite{francotte2006chirality}.
Numerous techniques have
been implemented to detect and discriminate opposite enantiomers with high precision.
These include now routine spectroscopies such as Circular Dichroism (CD)\cite{berova2000circular,he2011determination} or Optical Activity (OA)\cite{barron2009molecular} as well as more advanced ones like Raman Optical Activity (ROA)\cite{barron2009molecular} or Photoelectron CD (PECD)\cite{powis2008photoelectron}.
Differences of observables involving various polarization configurations permit to cancel the achiral contributions and single out the chiral ones.
Most chiral-sensitive spectroscopies suffer from an unfavorable signal-to-noise ratio since 
 the ratio of chiral to achiral signals is usually of the order of a percent or less (the ratio of molecular size to the optical wavelength).\\
The application of third-order nonlinear spectroscopies to measure chiral signals has gained experimental and theoretical interest over the past decades\cite{hache1999nonlinear,abramavicius2009coherent,
fischer2005nonlinear,mesnil2000experimental}.
Belkin and Shen\cite{belkin2005non} have  focused on second-order $\chi^{(2)}$ signals that vanish altogether in an achiral isotropic sample.
Most spectroscopic measurements of matter chirality are carried out on randomly oriented samples.
In non isotropic samples, important artefacts cover the CD signals such as linear dichroism or birefringence and must be dealt with\cite{castiglioni2009experimental}.
Here, we focus on molecules in the liquid or gas phase where the molecular response must be rotationally averaged.
Rotational averagings of cartesian\cite{craig1998molecular} and spherical\cite{andrews1982irreducible} tensors are well established.
We use the irreducible tensor formalism\cite{andrews1982irreducible,jerphagnon1978description} to carry out the rotational averages.
This is very convenient 
since only the $J=0$ tensor components do not vanish upon averaging.

In this study, we demonstrate that chiral nonlinear signals offer a way to control which chiral pathways contribute to the final signals by using various polarization and pulse geometry configurations.
In particular, we show that signals involving the chiral interaction at a given step along the interaction pathways can be extracted.
We have calculated chiral signals that only depend only on the magnetic dipole or on the electric quadrupole interactions. 
These allow to assign explicitly the multipolar nature of a given transition.
This is of importance to describe near field chiral interactions\cite{mun2019importance,rusak2019enhancement} and for the emergence of X-ray chiral sensitive signals where the relative magnitudes of electric quadrupole and magnetic dipole may be very different than at the visible and infrared frequency regimes\cite{yamamoto2008assignment}.

We focus on chiral four-wave-mixing (4WM) signals.
Several polarization schemes can single out chiral contributions by highlighting different types of interactions.
In section 2, we first present 4WM spectroscopies in general terms using the multipolar interaction Hamiltonian.
We compute all possible combinations of chiral-sensitive 4WM techniques and discuss the averaging of signals using the irreducible tensor representation of the 4WM response tensors.
Finally, in section \ref{sectionSibu}, we apply this formalism to study valence excitations in the drug S-ibuprofen by Stimulated X-ray Raman Spectroscopy (SXRS) and 2D Electronic Spectroscopy (2DES).

\section{Multipolar representation of four-wave-mixing signals}

We start with the multipolar radiation-matter coupling Hamiltonian that includes the electric and magnetic dipoles and the electric quadrupole:
\begin{equation}
H_\text{int}(t) = -\bm \mu\cdot \bold E(t)-\bm m\cdot \bold B(t)-\bm q\cdot \nabla\bold E(t)
\label{hintMULTI}
\end{equation}
We consider 3 four wave mixing techniques denoted $\bold k_I, \bold k_{II}$ and $\bold k_{III}$ according to their phase matching direction
\begin{eqnarray}
\bold k_I &=& - \bold k_1 +\bold k_2 + \bold k_3\\
\bold k_{II} &=& + \bold k_1 -\bold k_2 + \bold k_3\\
\bold k_{III} &=& + \bold k_1 +\bold k_2 - \bold k_3
\end{eqnarray}
$\bm k_1$, $\bm k_2$ and $\bm k_3$ are the wavevectors of the time-ordered incoming pulses. 
We use the vectors $(u_1,u_2,u_3) = (-1,1,1), \ (1,-1,1), \ (1,1,-1)$ to represent the $\bold k_I, \bold k_{II}$ and $\bold k_{III}$ techniques respectively.
In a three level system, the $\bold k_I$ and $\bold k_{II}$ techniques contain three pathways (excitated state emission ESE, ground state bleaching GSB and excited state absorption ESA) while $\bold k_{III}$ has only two ESA pathways.
These pathways indicate whether the molecule is back in the ground state after the first two interactions (GSB) or in an excited state (ESE and ESA)\cite{berera2009ultrafast}. 

The heterodyne-detected four-wave-mixing signal generally contains chiral and nonchiral components
\begin{equation}
S_{\text{het}}(\Gamma) = S_{\text{achir}}(\Gamma) + S_{\text{chir}}(\Gamma)
\end{equation}
\noindent where $\Gamma$ represents collectively the set of parameters that control the multidimensional signal (typically central frequencies, polarizations, bandwidths) as well as the wavevector configuration ($\bold k_I, \bold k_{II}$ or $\bold k_{III}$).
The achiral contribution $S_{\text{achir}}$ is given by the purely electric dipole contribution\cite{abramavicius2009coherent}:
\begin{multline}
S_{\text{achir}}(\Gamma) = -\frac{2}{\hbar} \Im\int dt dt_3 dt_2 dt_1 \bold \bold R_{\mu\mu\mu\mu}(t_3,t_2,t_1) \bullet (\bold E_s(t)\otimes \bold E_3(t-t_3) \otimes \bold E_2(t-t_3-t_2) \\
\otimes \bold E_1(t-t_3-t_2-t_1))
\label{heterodynenonchir}
\end{multline}
\noindent with 
\begin{equation}
\bold R_{\mu\mu\mu\mu}(t_3,t_2,t_1) = \Big(-\frac{i}{\hbar}\Big)^3 \langle \bm\mu_\text{left}\mathcal G(t_3)\bm\mu_-\mathcal G(t_2)\bm\mu_-\mathcal G(t_1)\bm\mu_-\rangle
\end{equation}
$\bold R_{\mu\mu\mu\mu}$ is a sum of pathways with four electric dipoles. 
At the lowest multipolar order, the chiral contribution $S_{\text{chiral}}$ contains either one magnetic dipole or one electric quadrupole and is given by 
\begin{eqnarray}
&& S_{\text{chiral}}(\Gamma) =  -\frac{2}{\hbar} \Im\int dt dt_3 dt_2 dt_1 \nonumber\\
&&( \bm R_{m\mu\mu\mu} \bullet (\bold B_s\otimes \bold E_3 \otimes \bold E_2 \otimes \bold E_1) + \bm R_{q\mu\mu\mu}\bullet ( \nabla \bold E_s\otimes \bold E_3 \otimes \bold E_2 \otimes \bold E_1)\nonumber\\
&&+ \bm R_{\mu m\mu\mu} \bullet (\bold E_s\otimes \bold B_3 \otimes \bold E_2 \otimes \bold E_1) + \bm R_{\mu q\mu\mu}\bullet (\bold  E_s\otimes \nabla \bold E_3 \otimes \bold E_2 \otimes \bold E_1)\nonumber\\
&&+\bm R_{\mu \mu m\mu} \bullet (\bold E_s\otimes \bold E_3 \otimes \bold B_2 \otimes \bold E_1) + \bm R_{\mu\mu q\mu}\bullet (\bold  E_s\otimes \bold E_3 \otimes\nabla \bold E_2 \otimes \bold E_1)\nonumber\\
&&+\bm R_{\mu\mu\mu m} \bullet (\bold E_s\otimes \bold E_3 \otimes \bold E_2 \otimes \bold B_1) + \bm R_{\mu\mu\mu q}\bullet (\bold E_s\otimes \bold E_3 \otimes \bold E_2 \otimes \nabla \bold E_1))
\label{chiraltot1}
\end{eqnarray}
\noindent where we have omitted the time variable for conciseness. The multipolar matter correlation functions are given by:\\
\begin{eqnarray}
\bold R_{m\mu\mu\mu}(t_3,t_2,t_1) &=& \Big(-\frac{i}{\hbar}\Big)^3 \langle \bm m_\text{left}\mathcal G(t_3)\bm\mu_-\mathcal G(t_2)\bm\mu_-\mathcal G(t_1)\bm\mu_-\rangle\\
\bold R_{q\mu\mu\mu}(t_3,t_2,t_1) &=& \Big(-\frac{i}{\hbar}\Big)^3 \langle \bm q_\text{left}\mathcal G(t_3)\bm\mu_-\mathcal G(t_2)\bm\mu_-\mathcal G(t_1)\bm\mu_-\rangle
\end{eqnarray}
The four magnetic dipole and the four electric quadrupole response functions in Eq. \ref{chiraltot1} are obtained by permuting the position of the $\bm m$ or $\bm q$ within each interaction pathway.
The subscripts $(\text{left})$  and $(-)$ indicates Liouville space superoperators\cite{harbola2008superoperator}. 
The Liouville space superoperators are defined by their action on Hilbert space operators as $\mathcal{O}_\text{left}\mathcal{A}\equiv\mathcal{O}\mathcal{A}$ and $\mathcal{O}_\text{right}\mathcal{A}\equiv\mathcal{A}\mathcal{O}$.  
We further define their linear combinations $\mathcal{O}_{\pm}$ which correspond to commutators and anticommutators in Hilbert space as $\mathcal{O}_{\pm}\mathcal{A}\equiv\mathcal{O}\mathcal{A}\pm\mathcal{A}\mathcal{O} .$
Such operators allow to keep track of interactions on the ket or bra side of the density matrix.

Assuming a slowly varying electric field envelope, we can express the magnetic fields and the electric field gradients as:
\begin{eqnarray}
\bm E_i(t) &=& \mathcal E_i(t) \ \bm \epsilon_i\\
\bm B_i(t) &=& \mathcal E_i(t) \ \frac{1}{c} u_i \bm{\hat k}_i \wedge \bm \epsilon_i\\
\bm \nabla E_i(t) &=& \mathcal E_i(t) \ i u_i \frac{\omega_i}{c}\bm{\hat k}_i \otimes \bm \epsilon_i
\end{eqnarray}
\noindent where $\bm \epsilon_i$ is the polarization unit vector of the $i$th pulse electric field.

The chiral contributions to 4WM are defined for rotationally-averaged samples and will be calculated using the irreducible tensor algebra\cite{jerphagnon1978description}. 
In this formalism, cartesian tensors are expanded in irreducible tensors, i.e. tensors transforming according to the irreducible representations of the rotation group SO(3): 
$T = \sum_{\tau J} \ _\tau T^J$ where $\tau$ is the seniority index which depends on the coupling scheme of matter quantities constituting the response tensor.
This formalism is a generalization of the decomposition of a matrix into its trace, an anti-symmetric part and a traceless symmetric part.
For 4WM signals, we apply it to rank 4 and 5 cartesian tensors.
Irreducible tensors up to irreducible rank $J=5$ appear in the decomposition of the matter and field tensors. The strength of the formalism resides in that only the isotropic $J=0$ tensors contributes to the rotationally averaged signals and thus need to be calculated.\cite{jerphagnon1978description,varshalovich1988quantum}
The signal is given by an irreducible tensor product of the matter response function $\bm R$ and the field tensor $\bm F$:
\begin{equation}
\bm R \cdot \bm F = \sum_{\tau J} \sum_{M=-J}^J (-1)^M\ _\tau R^{JM} \ _\tau F^{J-M}
\end{equation} 
The field tensor $F$ is kept general here and in practice will be described by a direct product of four field functions $\bm E, \bm B$ or $\nabla \bm E$. For example, the response tensor $R_{m\mu\mu\mu}$ and $R_{q\mu\mu\mu}$ are contracted with $\bm B \otimes\bm E\otimes\bm E\otimes\bm E$ and $\nabla \bm E \otimes\bm E\otimes\bm E\otimes\bm E$ respectively.

The rotationally-averaged contraction between matter and field response tensor is obtained by retaining only the $J=0$ terms:
\begin{equation}
(\bm R \cdot \bm F)_\text{av} = 
\sum_\tau \ _\tau R_\text{dip-E}^{J=0} \ _\tau F_\text{elec}^{J=0}
+\ _\tau R_\text{dip-M}^{J=0} \ _\tau F_\text{mag}^{J=0}
+\ _\tau R_\text{quad-E}^{J=0} \ _\tau G_\text{quad}^{J=0}
\end{equation}
where the general expression for the matter and the field tensors are given in Appendix A and in supplementary materials, section 1.
The field tensors $_\tau F_\text{elec}^{J=0}$ and $_\tau F_\text{mag}^{J=0}$ are calculated from the rank 4 cartesian tensors while the $_\tau G_\text{quad}^{J=0}$ is calculated from the rank 5 tensors involving $\nabla \bm E$.


As is done in circular dichroism (CD) or Raman optical activity (ROA), combinations of various polarization configurations can lead to the cancellation of the achiral electric dipole contribution and provides a measure of the chiral response function.
The electric dipole invariant tensor components have even parity while the magnetic dipole and electric quadrupole ones have an odd parity.

Unlike the linear CD, nonlinear 4WM signals offer multiple possible cancellation scenarios of the achiral components which can be combined in order to enhance desired features of the signal.
There are four possible chiral polarization configurations (denoted $\alpha, \beta, \gamma$ and $\delta$) for each of the phase matching directions ($\bold k_i = \bold k_{I}$, $\bold k_{II}$ and $\bold k_{III}$), leading to 12 possible schemes, see Eqs. \ref{chirtech1}-\ref{chirtech4}. 
Each of them further depends on the crossing angles of the incoming pulses.
There are thus many ways to access the chiral response functions with various degrees of control.

The four polarization schemes are given by
\begin{eqnarray}
S_\text{chir}(\alpha, \bm k_i) &=& S_\text{het}(L,L,L,L) - S_\text{het}(R,R,R,R)\label{chirtech1}\\
S_\text{chir}(\beta, \bm k_i) &=& S_\text{het}(L,R,L,R) - S_\text{het}(R,L,R,L)\\
S_\text{chir}(\gamma, \bm k_i) &=& S_\text{het}(L,L,R,R) - S_\text{het}(R,R,L,L)\\
S_\text{chir}(\delta, \bm k_i) &=& S_\text{het}(L,R,R,L) - S_\text{het}(R,L,L,R)\label{chirtech4}
\end{eqnarray}
\noindent where the arguments of $S_\text{het}(e_s,e_3,e_2,e_1)$ indicate the polarization of pulses $\bm E_s$, $\bm E_3$, $\bm E_2$ and $\bm E_1$ and $\bold k_i = \bold k_{I}, \bold k_{II}$ or $\bold k_{III}$.

We shall show that it is possible to experimentally select signals occurring solely through magnetic dipole or through electric quadrupole interactions. 
It is also possible to identify signals whereby the chiral interaction occurs at a given step in the interaction pathways.

\section{Resonant chiral SXRS and 2DES signals in S-ibuprofen}
\label{sectionSibu}
\subsection{Ibuprofen quantum chemistry}

\begin{figure*}[!h]
  \centering
  \includegraphics[width=0.5\textwidth]{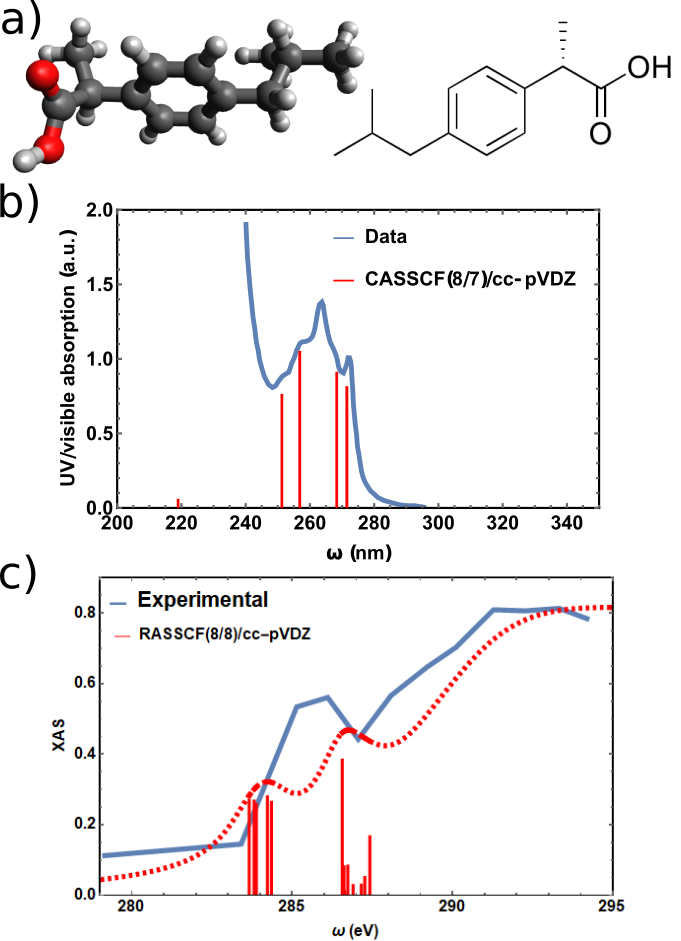}
  \caption{a) Molecular structure of S-ibuprofen. b) Experimental UV absorption from Moore et al.\cite{moore2002photophysical} (blue) and stick spectrum calculated at the cc-pVDZ/CASSCF(8/7) level. c) Experimental X-ray absorption \cite{riccardo2020} (blue) and stick spectrum at the cc-pVDZ/CASSCF(8/7). The dashed red curve is computed from the stick spectrum by convoluting with Lorentzian lineshapes and a step function accounting for ionization contributions. 
\label{fig2}}
\end{figure*}

The valence electronic excited states of S-ibuprofen (Fig. \ref{fig2}a) were computed using multi configurational self-consistent field (MCSCF) calculations at the cc-pVDZ/CASSCF(8/7) level of theory using the MOLPRO package\cite{werner2012molpro}.

The core-excited states were calculated at the cc-pVDZ/RASSCF(9/8) level by moving one by one the 1s carbon orbitals into the active space and freezing it to a single occupancy.
The second order Douglas-Kroll-Hess Hamiltonian was used to account for relativistic corrections\cite{reiher2006douglas}.
The valence and core excited states stick spectra compared with experiment\cite{moore2002photophysical,riccardo2020} are displayed in Fig. \ref{fig2} b) and c). 

\subsection{Chiral Stimulated X-ray Raman Spectroscopy}

\begin{figure*}[!h]
  \centering
  \includegraphics[width=0.5\textwidth]{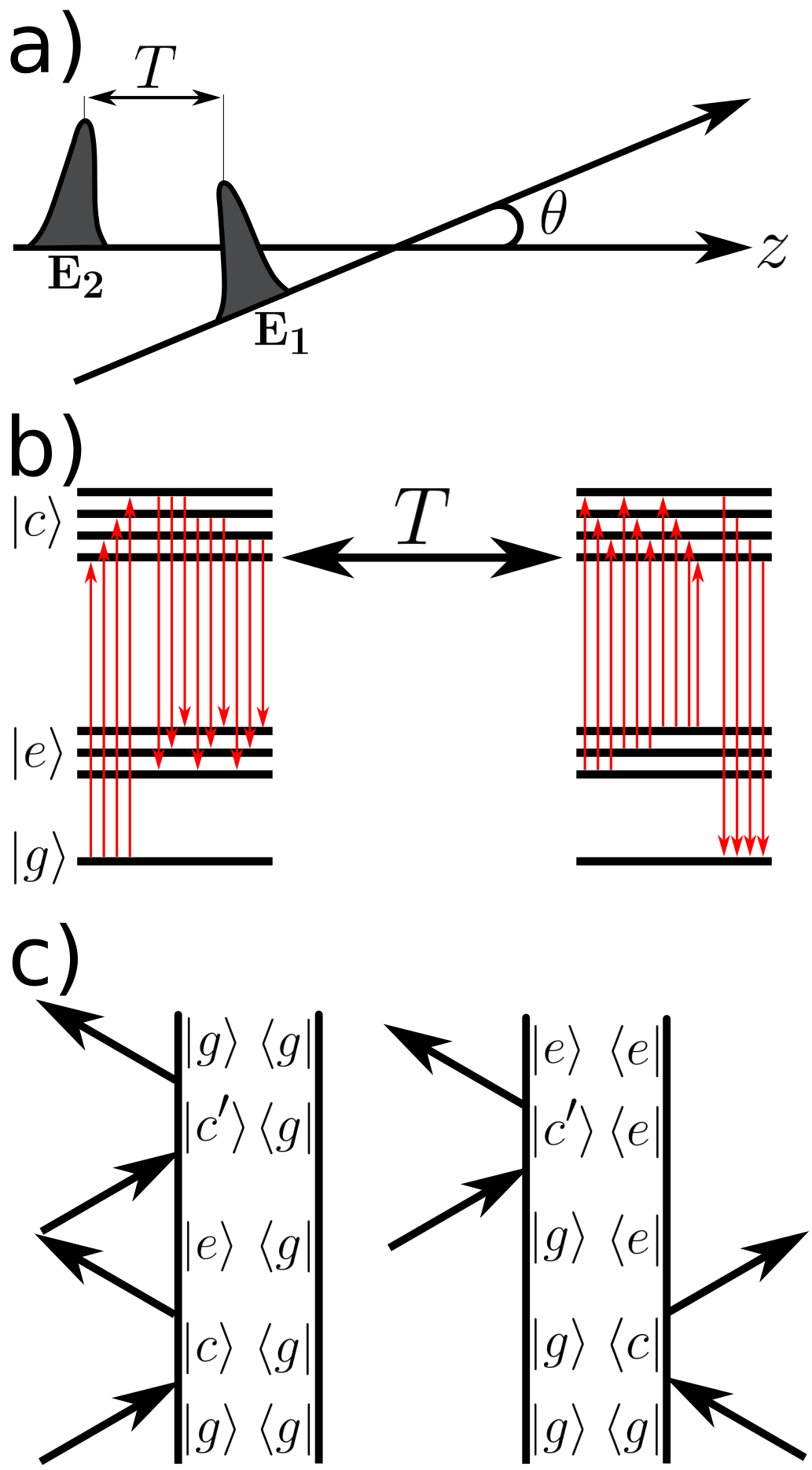}
  \caption{a) Pulse configuration for SXRS. b) Jablonski energy levels contributing to the SXRS signal. $g$ is the ground state, $e$ is the valence excited state manifold and $c$ the core excited state one. c) Ladder diagrams for SXRS. To account for the lack of time ordering of the interactions with the first pulse, each of these diagrams contributes twice to the signal. Upon Fourier transforming over the time delay, only the left diagram contributes to positive frequencies. 
\label{fig3}}
\end{figure*}

We now employ the polarization configurations developed earlier to compute the chiral Stimulated X-ray Raman Spectra (cSXRS), at the carbon K edge, as sketched in Fig. \ref{fig3}.
This pump-probe technique involves two ultrashort X-ray pulses, Fig. \ref{fig3}a, whose variation with their delay carries information on the valence excitation manifold. 
Each X-ray pulse induces a stimulated Raman process in the molecule (Fig. \ref{fig3}b) and its broad bandwidth allows to pump or probe many valence excited states in a single shot with high temporal resolution.
The core resonance allows to control which atoms are excited and probed, and the signal thus carries information about the valence excitations in the vicinity of the selected core.

The signal can be read off the loop diagrams displayed in Fig. \ref{fig3}c and reads
\begin{multline}
S_\text{SXRS}(T) = -\frac{2}{\hbar}\Im (\frac{i}{\hbar})^3
\int dt ds_3 ds_2 ds_1 \bm E_2^*(t)\bm E_2(t-s_3)\Big(\\
\langle \Psi(t_0)| G^\dagger(t)\bm  \mu G(s_3) \bm \mu^\dagger G(s_2) \bm \mu G(s_1) \bm \mu^\dagger G(t-s_3-s_2-s_1)|\Psi(t_0)\rangle  \bm E_1^*(t-s_3-s_2)E_1(t-s_3-s_2-s_1)\\
+
\langle \Psi(t_0)| G^\dagger(t-s_2-s_1)\bm  \mu G^\dagger(s_1) \bm \mu^\dagger G^\dagger(s_2) \bm \mu G(s_3) \bm \mu^\dagger G(t-s_3)|\Psi(t_0)\rangle \bm E_1(t-s_2) \bm E_1(t-s_2-s_1)\Big)
\end{multline}

By expanding this expression in molecular eigenstates, transforming the fields into the frequency domain and taking the Fourier transform over $T$, we obtain the following expressions for the electric dipole contribution to the signal:
\begin{equation}
S_\text{SXRS}(\Omega) = |E_2|^2|E_1|^2\frac{2}{\hbar^4}\Im\sum_{ecc'}
\bm \mu_{gc'}\bm \mu^\dagger_{c'e}\bm \mu_{ec}\bm \mu^\dagger_{cg}
\frac{I_{2,c'g}(\Omega)I_{1,cg}(\Omega)}{\Omega-\omega_{eg}+i\epsilon}
\label{eqSXRS_SOS}
\end{equation}
\begin{multline}
I_{2,c'g}(\Omega) = \frac{e^{-\Omega^2/4\sigma_2^2}}{2\sigma_2^2}e^{-z_2^2}(i+\text{erfi}(z_2)) \ \ \ \ \ \ \ \text{with} \ \ \ \ \ \ \ z_2 = -\frac{1}{\sigma_2}(\omega_2+\Omega/2-\omega_{c'g}+i \epsilon)\\
I_{1,cg}(\Omega) = \frac{e^{-\Omega^2/4\sigma_2^2}}{2\sigma_1^2}e^{-z_1^2}(i+\text{erfi}(z_1)) \ \ \ \ \ \ \ \text{with} \ \ \ \ \ \ \ z_1 = -\frac{1}{\sigma_1}(\omega_1+\Omega/2-\omega_{cg}+i \epsilon)
\end{multline}
\noindent where $I_{1,cg}(\Omega)$ and $I_{2,c'g}(\Omega))$ are the lineshape functions associated with the first and second X-ray pulses respectively.

\begin{figure*}[!h]
  \centering
  \includegraphics[width=0.5\textwidth]{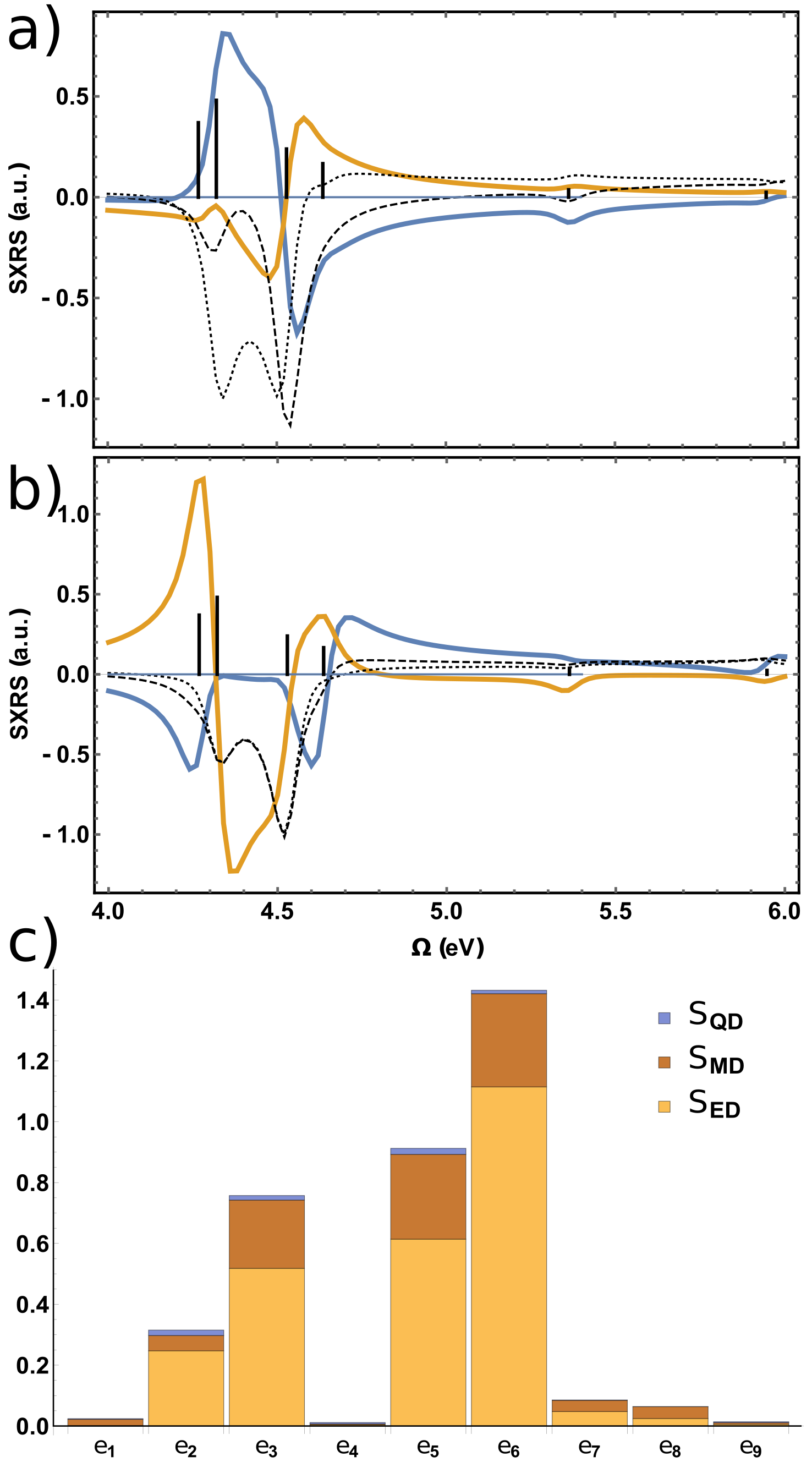}
  \caption{a) cSXRS($\Omega$,$\alpha$) signal of ibuprofen. Blue: both X-ray pulses are centred at the chiral carbon 1s core-excitation transition energy (283.8 eV). Orange: the first pulse is tuned at the chiral carbon 1s core transition and the second pulse is tuned at the carbon in the propyl group (286.6 eV). The dashed and dotted curves are the total SXRS signals with LL and RR polarization configurations respectively. The total SXRS signals are normalized and the cSXRS are multiplied by 5. In b) cSXRS($\Omega$,$\beta$) signal with similar pulse central frequencies. The dashed and dotted curves are the total SXRS signals with LR and RL polarization configurations respectively.
c) Relative multipolar contributions from each state to the total SXRS signals.
\label{fig4}}
\end{figure*}

The other contributions are obtained by replacing one of the electric dipoles in Eq. \ref{eqSXRS_SOS} with a magnetic dipole or with an electric quadrupole, following Eq. \ref{chiraltot1}.

SXRS is a $\bm k_\text{II}$ technique carried out with two non-collinear pulses.
This constrains the angles of the 4WM pulses defined above to $\theta_s = \theta_3 = 0$ and $\theta_2 = \theta_1 = \theta$.
Furthermore, the polarizations are not independent and $\bm e_s = \bm e_3$ and $\bm e_2 = \bm e_1$. We find the following two possible chiral signals
\begin{eqnarray}
S_\text{cSXRS}(\Omega,\alpha,\Gamma) &=& S_\text{cSXRS}(\Omega,LLLL) - S_\text{cSXRS}(\Omega,RRRR)\\ 
S_\text{cSXRS}(\Omega,\gamma,\Gamma) &=& S_\text{cSXRS}(\Omega,LLRR) - S_\text{cSXRS}(\Omega,RRLL)
\end{eqnarray}
\noindent $\Gamma$ denotes all control parameters $\{ \theta, \omega_1, \omega_2, \sigma_1, \sigma_2\}$ with $\theta$ the crossing angle between the two pulses, $\omega_1$ and $\omega_2$ are the pulses central frequencies and $\sigma_1$ and $\sigma_2$ their Gaussian envelope standard deviation.

\begin{figure*}[!h]
  \centering
  \includegraphics[width=0.5\textwidth]{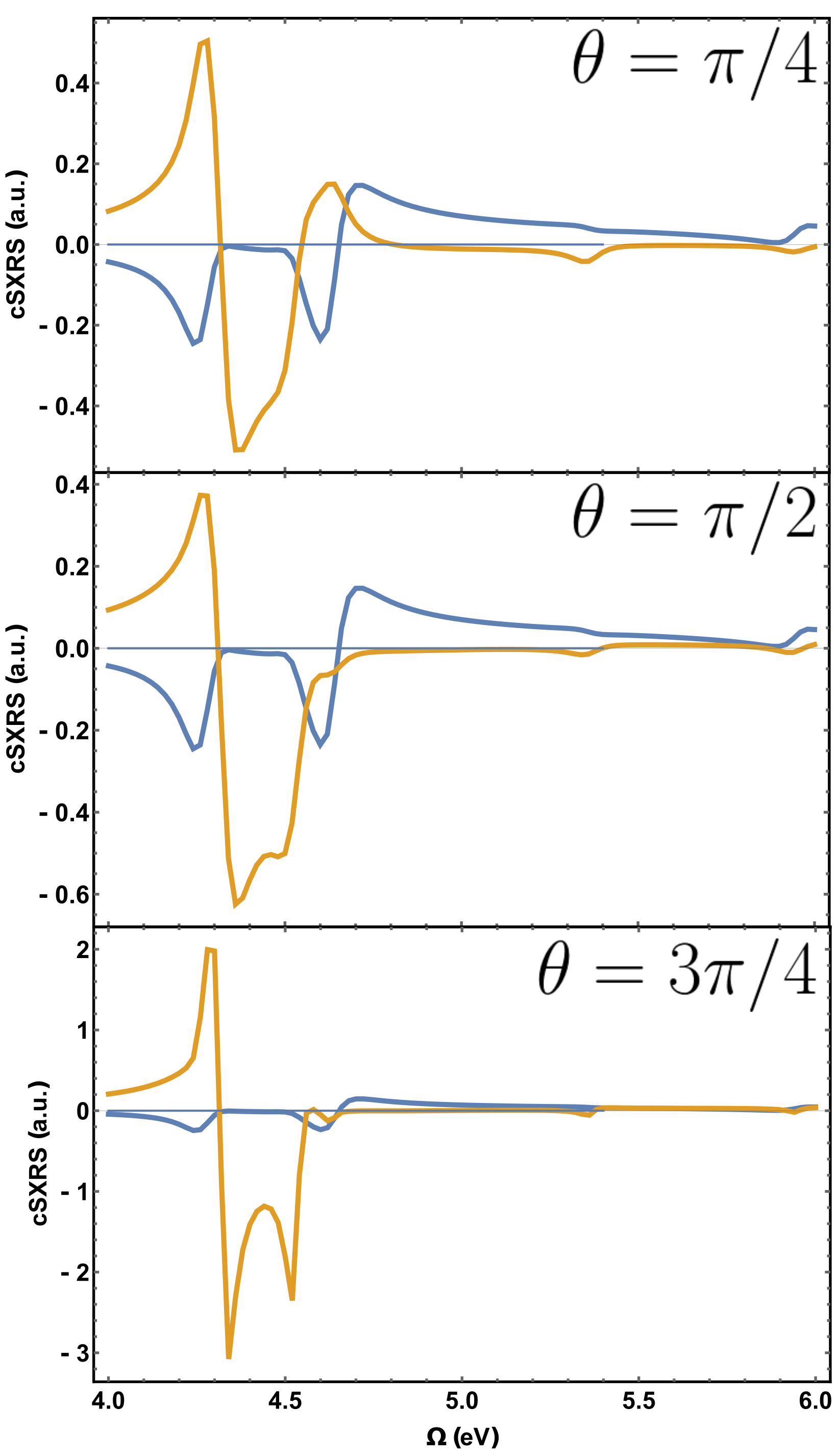}
  \caption{cSXRS($\Omega$,$\gamma$) for different crossing angles: $\pi/4$ (top), $\pi/2$ (middle) and $3\pi/4$ (bottom).
Central frequencies are the same as in Fig. \ref{fig4}.
\label{fig5}}
\end{figure*}

In Fig. \ref{fig4}a and b, we present the cSXRS spectra for the two polarization configurations $\alpha$ and $\gamma$ and a crossing angle $\theta = \pi / 4$.
In Fig. \ref{fig5}, we display the same signals for various pulses crossing angles.
Using the irreducible tensor formalism, we have calculated the contributions to the chiral signals for each polarization scheme and crossing angle.
Many interaction pathways are contributing for each signal and in Fig. \ref{fig4}c, we present the relative multipolar contribution of each state to the final signal at their resonant frequencies.
Other examples are in supplementary materials and a Mathematica code is provided to compute the contributions for any chosen configuration.


\begin{figure*}[!h]
  \centering
  \includegraphics[width=0.5\textwidth]{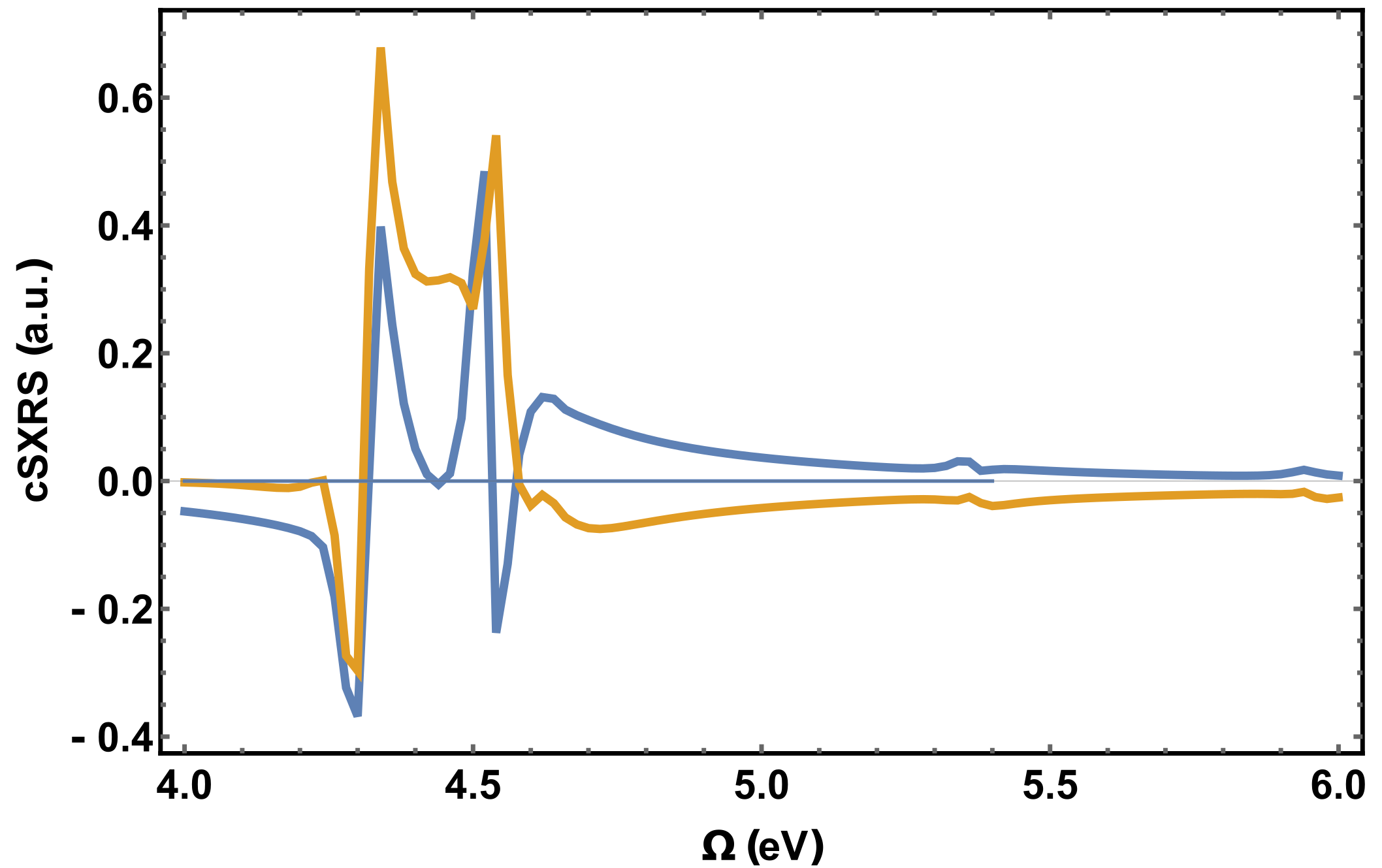}
  \caption{Sum of cSXRS signals with different crossing angles   extracting pathways with a chiral interaction only during the first pulse (in blue, Eq. \ref{combo1}) or during the second pulse (in orange, Eq. \ref{combo2}).
\label{fig6}}
\end{figure*}
Combining measurements with different crossing angles for the various chiral techniques ($\alpha$ or $\gamma$) allow to extract few selected contributions.
For example, Fig. \ref{fig6} shows SXRS signals in which the chiral interaction occurs only during the first pulse:
\begin{multline}
S_\text{cSXRS}(\Omega,\gamma,\frac{\pi}{4}) - S_\text{cSXRS}(\Omega,\gamma,3\frac{\pi}{4}) + S_\text{cSXRS}(\Omega,\alpha,\frac{\pi}{4}) - S_\text{cSXRS}(\Omega,\alpha,3\frac{\pi}{4}) =\\
-\Im\frac{2 i}{15} \big(5 \sqrt{6} \ _1R_{\mu \mu \mu m}-3 \sqrt{5} _1R_{\mu \mu \mu q}+5 \sqrt{6} _1R_{\mu \mu m\mu}\big)
\label{combo1}
\end{multline}
\noindent or during the second pulse:
\begin{multline}
S_\text{cSXRS}(\Omega,\gamma,\frac{\pi}{4}) - S_\text{cSXRS}(\Omega,\gamma,3\frac{\pi}{4}) - (S_\text{cSXRS}(\Omega,\alpha,\frac{\pi}{4}) - S_\text{cSXRS}(\Omega,\alpha,3\frac{\pi}{4})) =\\
-\Im\frac{2 i}{15} \big(5 \sqrt{6} \ _1R_{m\mu \mu \mu}+3 \sqrt{5} _1R_{q\mu \mu \mu}+5 \sqrt{6} _1R_{\mu m\mu\mu}\big)
\label{combo2}
\end{multline}


\subsection{Chiral 2D Electronic Spectroscopy}

We next apply the 4WM polarization schemes to 2DES of S-ibuprofen.
2DES has been used to resolve excitonic couplings and energy transfer in molecular aggregates.
It can further separate homogeneous and inhomogeneous contributions to absorption lineshapes.
Here, we apply it to S-ibuprofen using phenomenological relaxation and  compare with the SXRS signals that also probe the valence excited manifold.
2DES lacks the element-sensitivity of the X-rays but offers additional polarization and crossing angle controls because each interaction corresponds to a different pulse.

\begin{figure*}[!h]
  \centering
  \includegraphics[width=0.7\textwidth]{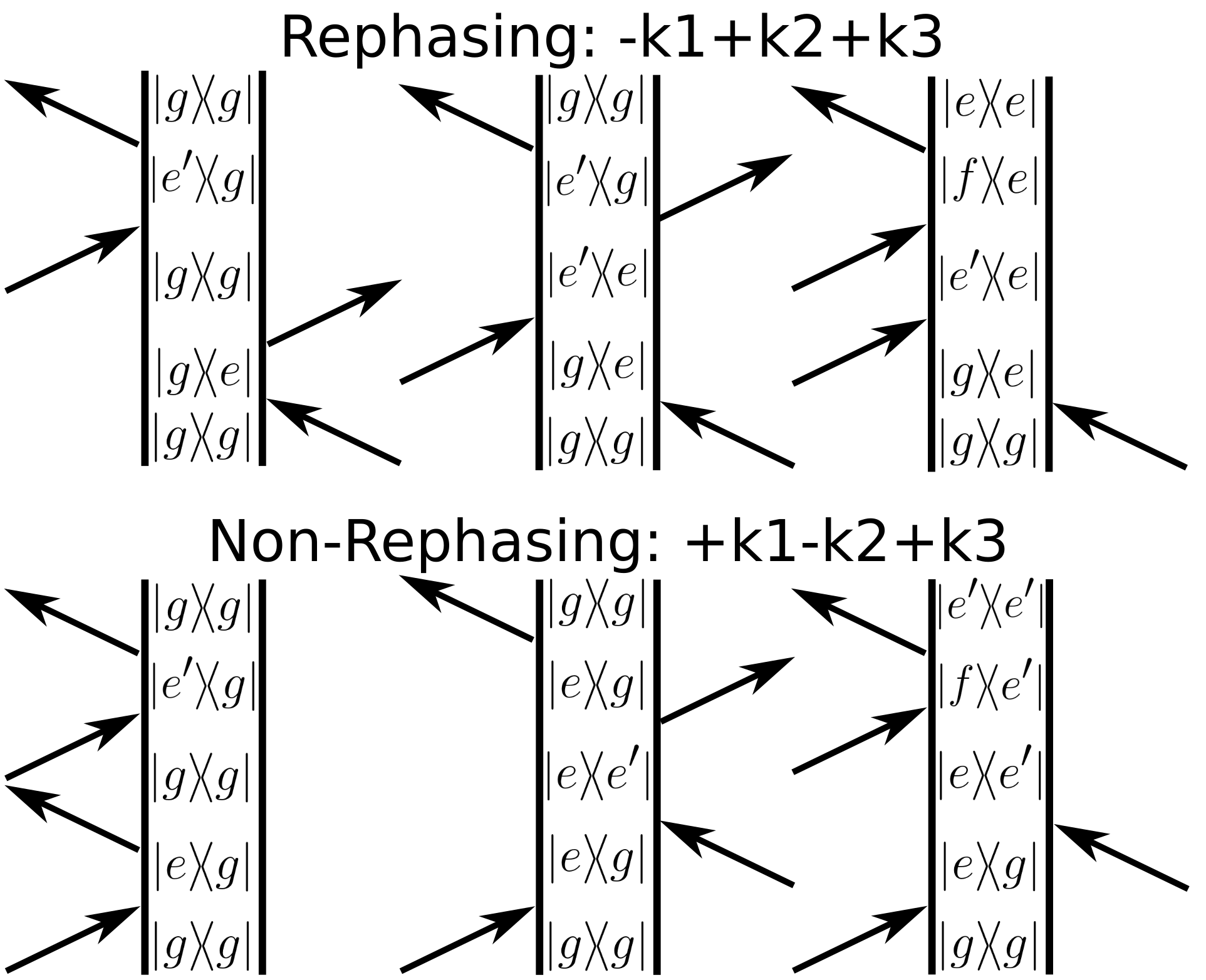}
  \caption{Ladder diagrams for 2DES.
\label{loop2DES}}
\end{figure*}
We focus on 2DES non-rephasing ($\bold k_{II}$ technique) signals whose diagrams are given in Fig. \ref{loop2DES}.
The correlation functions and SOS expressions of the 2DES signals are given in supplementary materials.

The freedom to independently select the polarization scheme and crossing angle for each of the four interactions results in a huge number of possible techniques.
Here, we focus on a single case and other combinations can be computed using the supplementary materials.
The chiral 2DES signal can be made to be sensitive to the electric quadrupole interaction only.
\begin{multline}
S_\text{c2DES}(\Omega_1,T_2,\Omega_3,\alpha) = 
S_\text{2DES}(\Omega_1,T_2,\Omega_3,LLLL) - S_\text{2DES}(\Omega_1,T_2,\Omega_3,RRRR)
\label{eq26}
\end{multline}

Using $,\theta_1=\pi/2,\theta_2=-\pi/2$ and $\theta_3=\pi/2$ as crossing angles, the pathways that contributes to the final signal are:
\begin{equation}
S_\text{c2DES}(\Omega_1,T_2,\Omega_3,\alpha) = 
\frac{3 \ _1 R_{\mu q\mu \mu}-\sqrt{3} \  _2 R_{\mu q\mu \mu}}{12 \sqrt{10}}
\end{equation}

\begin{figure*}[!h]
  \centering
  \includegraphics[width=1.\textwidth]{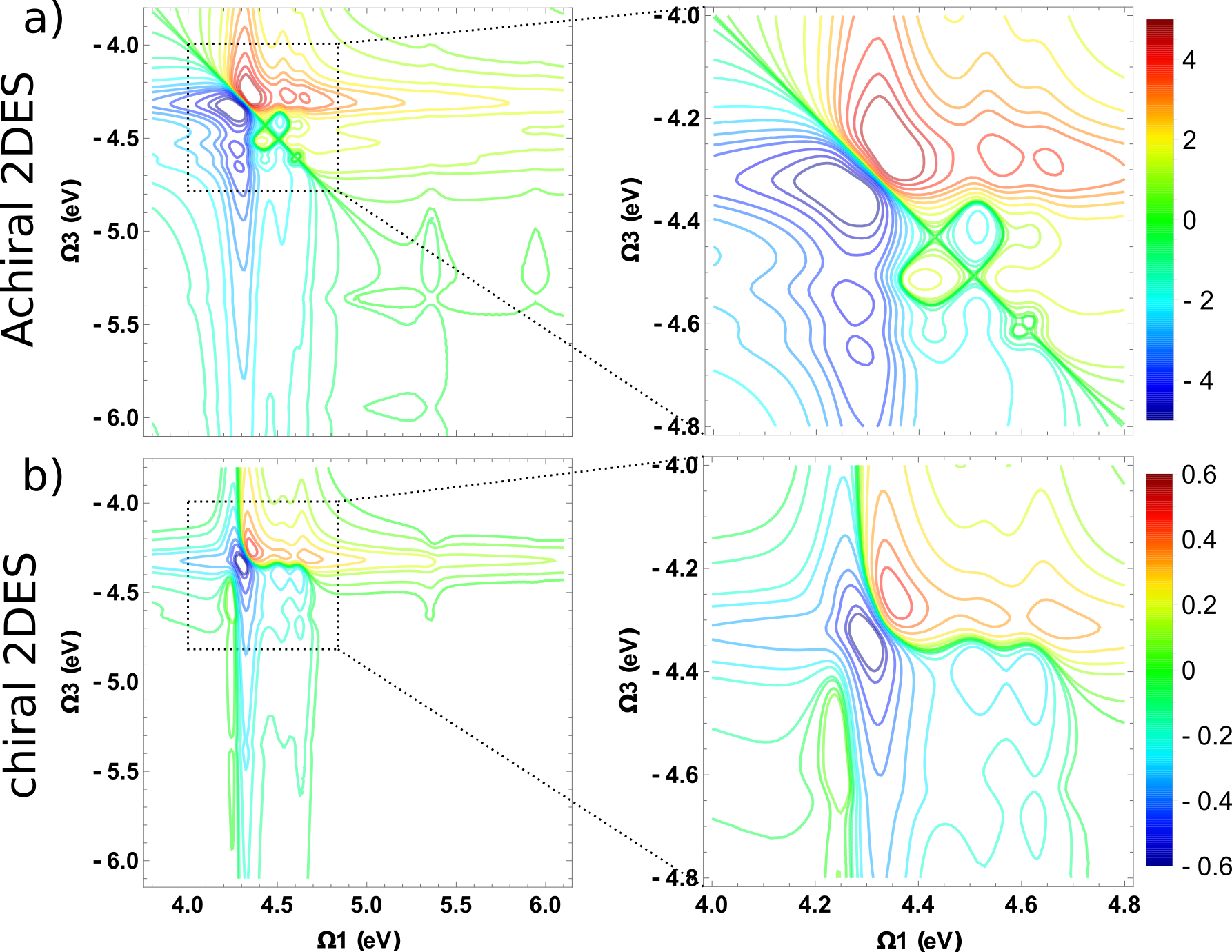}
  \caption{a) Achiral 2DES spectrum on an isotropic average (Eq. XX) of S-ibuprofen calculated in the dipolar approximation. b) Chiral 2DES defined in Eq. \ref{eq26} with crossing angles $\theta_1=\pi/2,\theta_2=-\pi/2$ and $\theta_3=\pi/2$. This choice of polarization and crossing angles singles out only electric quadrupole interaction in the chiral interactions. The right panels show the main features on an expanded scale.
\label{fig2DES}}
\end{figure*}

Fig.\ref{fig2DES} compares the electric dipole contribution (panels a and b) with the quadrupole-sensitive c2DES signal.
The dominant states in the achiral contribution (at 4.5 eV) are differnt from the ones in the chiral electric quadrupole one (at 4.3 eV).
In SXRS, the 4.3 and 4.53 eV transitions were giving the main contributions to the signals. 
On the other hand, in the chiral 2DES signal displayed in Fig. \ref{fig2DES} which contains only quadrupolar interactions, the 4.53 eV is contributing weakly.

\section{Conclusions}

Chiral 4WM signals can combine the detailed information of 4WM techniques on molecular dynamics and relaxation with the inherent structural aspects of chirality.
Their interpretation based on the multipolar expansion truncated at the magnetic dipole and electric quadrupole order is complicated by the many possible interaction pathways.
We have used the irreducible tensor formalism to calculate each contribution to the signals for various pulse configurations and circularly polarized light.

Applications are made to two chiral-sensitive 4WM techniques, SXRS and 2DES. 
While the former is more restricted since multiple interactions occur with the same pulse, it allows to add the extra element-sensitivity of the X-rays.
2DES on the other hand uses optical or UV pulses but is a well-established tabletop technique that has better control of the crossing angles and pulses polarizations.

Some polarization configurations are identified which permit to isolate specific contributions in which the chiral interaction (with the magnetic dipole or with the electric quadrupole) occurs at a chosen step within the interaction pathways. 
We further have shown how to extract chiral contributions involving only the magnetic dipole or the electric quadrupole.

\section*{Acknowledgements}

We acknowledge support of the National Science Foundation
(grant CHE-1953045) and of the Chemical
Sciences, Geosciences, and Biosciences division, Office of Basic
Energy Sciences, Office of Science, U.S. Department of Energy,
through award No. DE-FG02-04ER15571.
J.R.R. was supported by the DOE grant. Support for A.R. was provided by NSF via grant number CHE-1663822.

\begin{appendices}

\section{Chiral-sensitive 4WM signals}

$_\tau\bm R_\text{dip-E}^{J=0}$ and $_\tau\bm R_\text{dip-M}^{J=0}$ are rank 4 tensors that have the following rotational invariants:
\begin{eqnarray}
_0 R^{J=0} &=& \{\{\bm A\otimes \bm B\}_0 \otimes \{\bm C\otimes \bm D\}_0\}_0 = \frac{1}{3}(\bm A\cdot \bm B)(\bm C\cdot \bm D) \\
_1 R^{J=0} &=& \{\{\bm A\otimes \bm B\}_1 \otimes \{\bm C\otimes \bm D\}_1\}_0 = \frac{1}{\sqrt 3}(\bm A\wedge \bm B)\cdot(\bm C\wedge \bm D)\\
_2 R^{J=0} &=& \{\{\bm A\otimes \bm B\}_2 \otimes \{\bm C\otimes \bm D\}_2\}_0 \\ 
&=& \frac{1}{\sqrt 5}(\frac{1}{2}(\bm A\cdot \bm C)(\bm B\cdot \bm D)-\frac{1}{3}(\bm A\cdot \bm B)(\bm C\cdot \bm D)+\frac{1}{2}(\bm A\cdot \bm D)(\bm B\cdot \bm C))
\end{eqnarray}
\noindent where $A, B, C$ and $D$ can be either an electric dipole $\bm \mu$ or a magnetic dipole $\bm m$ interaction.
Each of these invariants gets contracted with the corresponding field invariants $_\tau F^{J=0}$:
\begin{eqnarray}
_0 F(\bm p_s,\bm p_3,\bm p_2,\bm p_1)^{J=0} &=& \{\{\bm p_s\otimes \bm p_3\}_0 \otimes \{\bm p_2\otimes \bm p_1\}_0\}_0 \\
_1 F(\bm p_s,\bm p_3,\bm p_2,\bm p_1)^{J=0} &=& \{\{\bm p_s\otimes \bm p_3\}_1 \otimes \{\bm p_2\otimes \bm p_1\}_1\}_0)\\
_2 F(\bm p_s,\bm p_3,\bm p_2,\bm p_1)^{J=0} &=& \{\{\bm p_s\otimes \bm p_3\}_2 \otimes \{\bm p_2\otimes \bm p_1\}_2\}_0 
\end{eqnarray}
\noindent where $\bm p_i = \bm \epsilon_i$ or $\frac{i}{c} \bm{\hat k}_i \wedge \bm \epsilon_i$.
$_\tau R_\text{quad-E}^{J=0}$ is a rank 5 tensor that has two rotational invariants:
\begin{eqnarray}
_0  R_\text{quad-E}^{J=0} &=& \{\{\bm \mu \otimes \bm \mu\}_1 \otimes \{\bm \mu\otimes \bm q\}_1\}_0 \\
_1  R_\text{quad-E}^{J=0} &=&\{\{\bm \mu \otimes \bm \mu\}_2 \otimes \{\bm \mu\otimes \bm q\}_2\}_0 
\end{eqnarray}
The relevant field tensors are:
\begin{eqnarray}
_0 G(\bm a,\bm b,\bm c,i\bm k\otimes d)^{J=0} &=& 
\{\{\bm a\otimes \bm b\}_1 \otimes \{\bm c\otimes \{i\bm k\otimes \bm d\}_2\}_1\}_0 \\
_1 G(\bm a,\bm b,\bm c,i\bm k\otimes d)^{J=0} &=&
\{\{\bm a\otimes \bm b\}_2 \otimes \{\bm c\otimes \{i\bm k\otimes \bm d\}_2\}_2\}_0
\end{eqnarray}
\noindent where $\bm a,\bm b,\bm c,\bm d = \bm \epsilon_i$.
We assume that the heterodyning pulse propagates along $z$ (hence $\bm{\hat k}_s = \bm e_0$). 
The three exciting pulses are non-collinear, forming an angle $\theta_i$ with the $y$ axis. 
This corresponds to the most common experimental situation where multiple pulses parallel to the optical table are incident on the sample with a small angle between them.
We further assume that the electric field polarization $\bm\epsilon_1, \bm\epsilon_2, \bm\epsilon_3$ are left or right polarized in the laboratory frame. 
In the irreducible basis $\{\bm e_1, \bm e_0, \bm e_{-1}\}$, the polarization vectors are given by
\begin{equation}
\label{basisdef}
\bm e_L = \begin{pmatrix}
1 \\ 
0 \\ 
0
\end{pmatrix} 
\ \ \ \ \ \ \ 
\bm e_R =\begin{pmatrix}
0 \\ 
0 \\ 
1
\end{pmatrix} 
\end{equation}
\noindent where $\bm e_L$ and $\bm e_R$ are the left and right-handed polarization vectors for a plane wave propagating along $z$.
For non-collinear pulses, the Wigner $\mathcal D^{J=1}$ matrix can be used to obtain the corresponding left and right polarizations.
The vectors $(\bm e_L, \bm e_R, \bm e_z)$ form an orthonormal basis with the following properties:
\begin{eqnarray}
&&\bm e_L^* = - \bm e_R \hspace{2.4cm} \bm e_R^* = - \bm e_L\nonumber\\
&&\bm e_L\cdot \bm e_L = \bm e_R\cdot \bm e_R = 0 \hspace{0.5cm} \bm e_L\cdot \bm e_R = -1\nonumber\label{basisdotprod}\\
&&\bm e_z \times \bm e_L = - i \bm e_L \hspace{1.4cm} \bm e_z \times \bm e_R =  i \bm e_R\label{basiscrossprod}
\end{eqnarray}
Eq. \ref{basiscrossprod} allows to get the polarization vectors $\bm b$ of the magnetic field 
\begin{equation}
\bm b_{L/R} = \frac{1}{c} \bm e_z \times \bm e_{L/R}
\end{equation}

In supplementary materials, we give the non-vanishing tensor components corresponding to different polarization schemes and the general tensor expressions for arbitrary polarization vectors in term of Clebsch-Gordan sums.

\end{appendices}

\section{Supporting Information}

Details on the quantum chemistry of S-ibuprofen and on the 2DES and SXRS signal derivations are given in supplementary materials.
The program allowing to extract the response tensor irreducible components contributing to a given configuration is also available.

\bibliographystyle{achemso}
\bibliography{c4WM}

\newpage

\begin{figure*}[!h]
  \centering
  \includegraphics[width=0.8\textwidth]{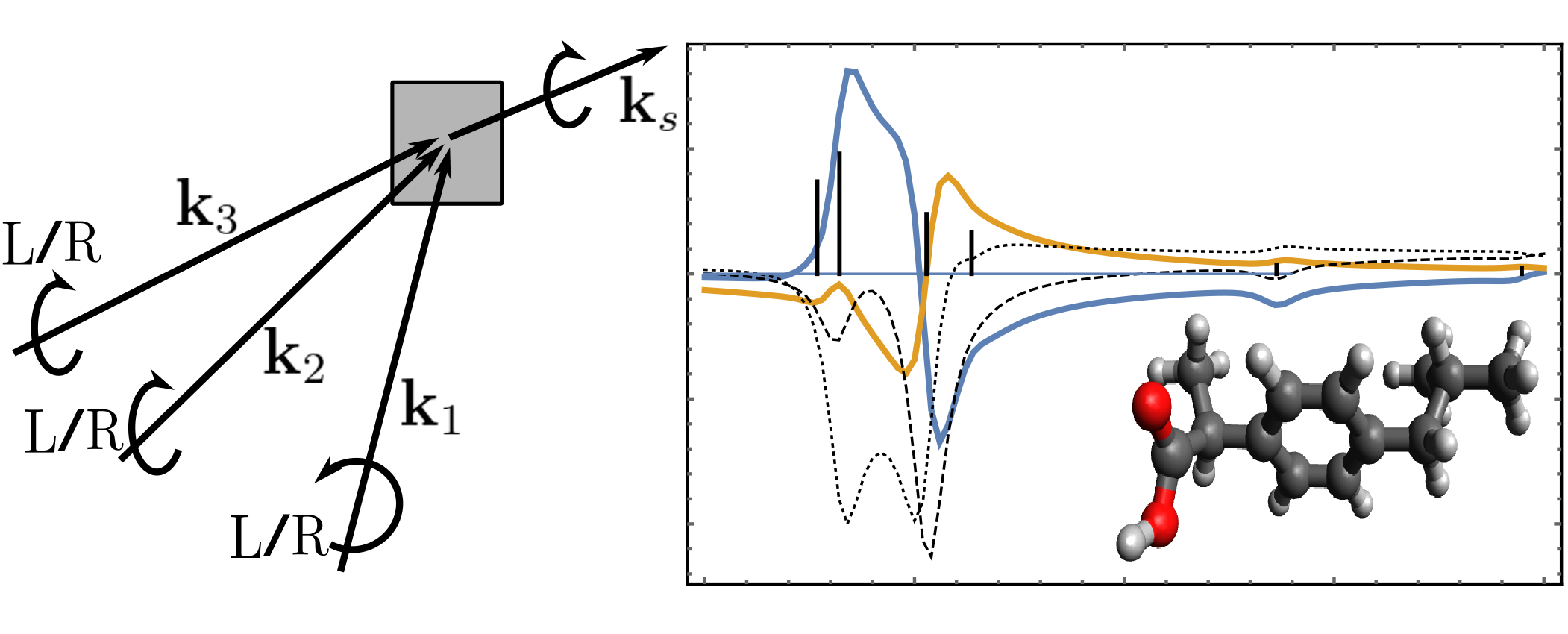}
  \caption{TOC graphics
\label{figTOC}}
\end{figure*}

\end{document}


\maketitle

\section{Non-colinear pulse configuration}
\label{appTENSGEN}

General expression for the rotationally invariant irreducible tensor components of the rank 4 and rank 5 field tensors invariants are provided explicitly in this section using the definition of the irreducible tensor product: $\{A \otimes B\}^{JM} = \sum_{m_A} C^{JM}_{j_A m_A j_B M-m_A} A^{j_A m_A}B^{j_B M-m_A}$ where $C^{JM}_{j_A m_A j_B M-m_A}$ is a Clebsch-Gordan coefficient.\\

\noindent\textbf{Electric dipole or magnetic dipole tensor components}
\begin{equation}
_0 F(\bm p_s,\bm p_3,\bm p_2,\bm p_1)^{J=0} = \sum_{ij} C_{1i1-i}^{00}C_{1j1-j}^{00}
[\bm p_s]^i[\bm p_3]^{-i}[\bm p_2]^j[\bm p_1]^{-j}
\end{equation}

\begin{equation}
_1 F(\bm p_s,\bm p_3,\bm p_2,\bm p_1)^{J=0} = \sum_{ijk} 
C_{1i1-i}^{00} C_{1j1i-j}^{1i} C_{1k1-i-k}^{1-i}
[\bm p_s]^{j}[\bm p_3]^{i-j}[\bm p_2]^{k}[\bm p_1]^{-i-k}
\end{equation}

\begin{equation}
_2 F(\bm p_s,\bm p_3,\bm p_2,\bm p_1)^{J=0} = \sum_{ijk} 
C_{2i2-i}^{00} C_{1j1i-j}^{2i} C_{1k1-i-k}^{2-i}
[\bm p_s]^{j}[\bm p_3]^{i-j}[\bm p_2]^{k}[\bm p_1]^{-i-k}
\end{equation}

\noindent\textbf{Electric quadrupole tensor components}

\begin{equation}
_0 G(a,b,c,d\otimes e)^{J=0} = \sum_{ijkl} 
C_{1i1-i}^{00}C_{1j1i-j}^{1i}C_{1k2-i-k}^{1-i}C_{1l1-i-k-l} 
[a]^{j}[b]^{i-j}[c]^{k}[d]^l[e]^{-i-k-l}
\label{quad1}
\end{equation}
\begin{equation}
_1 G(a,b,c,d\otimes e)^{J=0} = \sum_{ijkl} 
C_{2i2-i}^{00}C_{1j1i-j}^{2i}C_{1k2-i-k}^{2-i}C_{1l1-i-k-l} 
[a]^{j}[b]^{i-j}[c]^{k}[d]^l[e]^{-i-k-l}
\label{quad2}
\end{equation}

The $d\otimes e$ represents the $\bm k \otimes \bm \epsilon$ is the main text. Since the electric quadrupole is a $J=2$ irreducible tensor by definition, only the $J=2$ component of $d\otimes e$ enters into the irreducible tensor products given above.
Eqs. \ref{quad1} and \ref{quad2} are used to compute the field tensors that get contracted with the $J=0$ components of $R_{\bm \mu \bm \mu \bm \mu q}$. Similarly, $_{0/1} G(a,b,d\otimes e,c)^{J=0}$, $_{0/1} G(a,d\otimes e,b,c)^{J=0}$ and $_{0/1} G(d\otimes e,a,b,c)^{J=0}$ are defined for contractions with $R_{\bm \mu \bm \mu q \bm \mu}$, $R_{\bm \mu  q \bm \mu \bm \mu}$ and $R_{q \bm \mu \bm \mu \bm \mu }$ respectively.

\section{Quantum chemistry details}

Molecular orbitals included in the active space of the the cc-pVDZ/CASSCF(8/7) valence excited state computation are displayed in Fig. \ref{figS1}.
The ground state and the first 9 excited states were computed using this active space. 
There relative energies from the ground state energy was obtained as:
(0.0484956, 0.0604728, 0.10202, 0.167855, 0.16981, 0.177436, 0.181381, 0.208095, 0.229564) atomic units.
A global energy shift of 0.167 au to match the experimental spectrum was applied. Details of the molpro code can be shared upon reasonable request.

\begin{figure*}[!h]
  \centering
  \includegraphics[width=1.\textwidth]{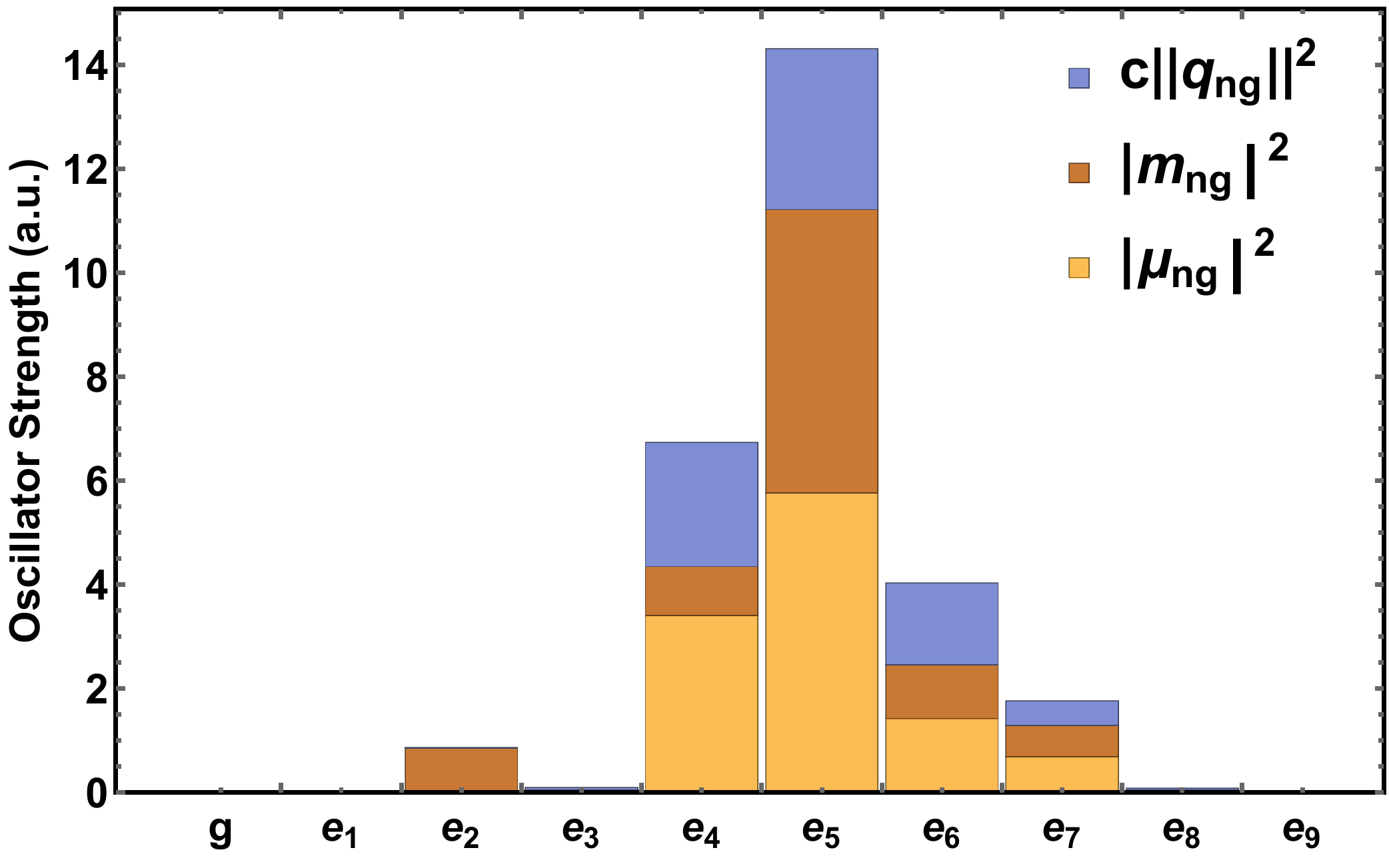}
  \caption{Norms of the transition multipoles matrix elements from the ground state to each of the calculated excited states.
\label{figS2}}
\end{figure*}

\begin{figure*}[!h]
  \centering
  \includegraphics[width=1.\textwidth]{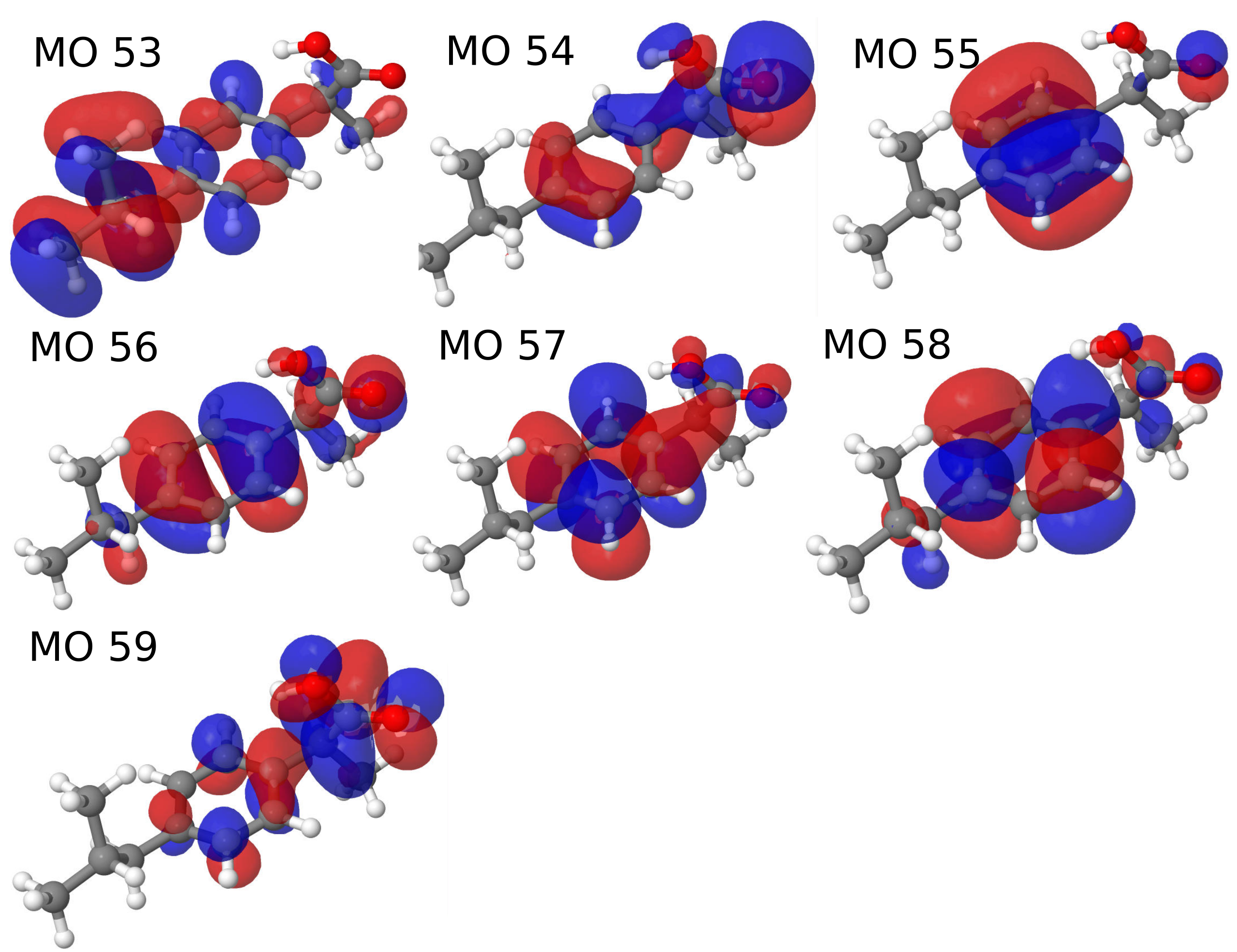}
  \caption{Molecular orbitals included in the active space.
  The HOMO orbital is MO56 and the LUMO one is MO57.
\label{figS1}}
\end{figure*}

The norm of the multipolar transition matrix elements is displayed in Fig. \ref{figS2}.
The states $e_1$, $e_2$ and $e_3$ have neglectable electric dipole transition matrix elements are thus dark states.
These states are not discussed further in this study.\\

Main Configuration State Functions (CSF) for each states are now given:

\textbf{Ground state}

 2222000:           0.9494771\\          

\textbf{State $e_1$}

 222/$\backslash$ 00:           0.6601039
 \hspace{2cm}
 22/20$\backslash$0:          -0.5920438\\

\textbf{State $e_2$}

 2/2200$\backslash$:           0.7762152
 \hspace{2cm} 
 222/00$\backslash$:          -0.3304313
 \hspace{2cm} 
 2/22$\backslash$00:          -0.2996679\\

\textbf{State $e_3$}

 222/0$\backslash$0:           0.6509050
 \hspace{2cm} 
 22/2$\backslash$00:           0.6238821\\

\textbf{State $e_4$}

 22/20$\backslash$0:           0.5071633
 \hspace{2cm} 
 222/$\backslash$00:           0.4184930
 \hspace{2cm} 
 /222$\backslash$00:          -0.4079125
 
 222/0$\backslash$0:           0.2354199
 \hspace{2cm} 
 2/22$\backslash$00:           0.2305255\\

\textbf{State $e_5$}

 22/2$\backslash$00           0.6272697
 \hspace{2cm} 
 222/0$\backslash$0          -0.5673681\\

\textbf{State $e_6$}

 /222$\backslash$00:           0.4715856
 \hspace{2cm} 
 /2220$\backslash$0:           0.3697852
 \hspace{2cm} 
 222/$\backslash$00:           0.3631566
 
 22/20$\backslash$0:           0.3484628
 \hspace{2cm} 
 22/$\backslash$/$\backslash$0:           0.2534831
 \hspace{2cm} 
 22/$\backslash$200:           0.2309925\\

\textbf{State $e_7$}

 /2220$\backslash$0:           0.5949174
 \hspace{2cm} 
 /222$\backslash$00:          -0.3073049
 \hspace{2cm} 
 22/$\backslash$020:           0.2481364
 
 2220020:           0.2300734
 \hspace{2cm} 
 2220/$\backslash$0:          -0.2286009
 \hspace{2cm} 
 22/20$\backslash$0:          -0.2151185
 
 222/$\backslash$00:          -0.2016753\\

\textbf{State $e_8$}

 2/2$\backslash$0/$\backslash$:          -0.3961695
 \hspace{2cm} 
 2/220$\backslash$0:           0.3510396
 \hspace{2cm} 
 2/2/0$\backslash$ $\backslash$:          -0.3189000
 
 22200/$\backslash$:          -0.2875625
 \hspace{2cm} 
 222/00$\backslash$:           0.2738819
 \hspace{2cm} 
 2/$\backslash$2/0$\backslash$:          -0.2581666\\

\textbf{State $e_9$}

 2/22$\backslash$00:           0.4754603
 \hspace{2cm} 
 2/2$\backslash$/0$\backslash$:          -0.3402749
 \hspace{2cm} 
 2/2/$\backslash$0$\backslash$:          -0.2871956
 
 2/$\backslash$20/$\backslash$:          -0.2778902
 \hspace{2cm} 
 2220/0$\backslash$:          -0.2379281\\

\section{SXRS integrals calculation}

Reading out the SXRS diagrams and summing over electronic states gives:
\begin{multline}
S_\text{SXRS}(\Omega) = \frac{2}{\hbar^4}\Im\sum_{ecc'}\int d\omega_2 d\omega_1 \Big(
\frac{\bm E_2^*(\omega_2)\bm E_2(\omega_2-\Omega)\bm E_1^*(\omega_1)\bm E_1(\omega_1+\Omega)
\bm \mu_{gc'}\bm \mu^\dagger_{c'e}\bm \mu_{ec}\bm \mu^\dagger_{cg}}
{(\omega_2-\omega_{c'g}+i\epsilon)(\Omega-\omega_{eg}+i\epsilon)(\omega_1+\Omega-\omega_{cg}+i\epsilon)} \\
+
\frac{\bm E_2^*(\omega_2)\bm E_2(\omega_2-\Omega)\bm E_1(\omega_1)\bm E_1^*(\omega_1-\Omega)
\bm \mu_{ec'}\bm \mu^\dagger_{c'g}\bm \mu^\dagger_{ec}\bm \mu_{cg}}
{(\omega_2-\Omega-\omega_{c'g}+i\epsilon)(\Omega+\omega_{eg}+i\epsilon)(-\omega_1+\Omega+\omega_{cg}+i\epsilon)}
\Big) 
\end{multline}

In the following, we specialize only in the first term that gives the positive frequency of the signal, and the goal is to calculate the following integral:
\begin{equation}
I_2(\Omega) = |E_2|^2\int_{-\infty}^{+\infty} d\omega_2
\frac{\bm E_2^*(\omega_2)\bm E_2(\omega_2-\Omega)}{
\omega_2-\omega_{c'g}+i\epsilon}
\end{equation}

Defining $\omega_c$ and $\sigma_2$ as the normalized Gaussian envelope parameter the integral, we have:
\begin{equation}
I_2(\Omega) = |E_2|^2\int_{-\infty}^{+\infty} d\omega_2
\frac{(
1/\sqrt{2\pi}\sigma_2)e^{-(\omega_2-\omega_c)^2/2\sigma_2^2}(1/\sqrt{2\pi}\sigma_2)e^{-(\omega_2-\omega_c-\Omega)^2/2\sigma_2^2}
}{
\omega_2-\omega_t+i\epsilon}
\end{equation}

\subsection{Integral calculation}

We first calculate
\begin{equation}
I = \int_{-\infty}^{+\infty} d x \frac{e^{-x^2}}{x-x_0}
= \int_{-\infty}^{+\infty} d x (x+x_0)\frac{e^{-x^2}}{x^2-x_0^2} 
= x_0 \int_{-\infty}^{+\infty} d x \frac{e^{-x^2}}{x^2-x_0^2}
= 2x_0 \int_{0}^{+\infty} d x \frac{e^{-x^2}}{x^2-x_0^2}
\end{equation}

We now define 
\begin{equation}
I(a) = 2x_0 \int_{0}^{+\infty} d x \frac{e^{-ax^2}}{x^2-x_0^2}
\end{equation}
The integral that we want to calculate is $I(1)$.
By deriving over $a$, we can write down the following equation on $I(a)$:
\begin{equation}
\frac{\partial}{\partial a} I(a) + z_0^2 I(a) = -z_0 \sqrt{\frac{\pi}{a}}
\end{equation}

Solving the ODE gives
\begin{equation}
I(a) = I(0) e^{-az_0^2} - 2\sqrt{\pi}e^{-az_0^2}\int_0^{\sqrt{a}z_0} dz \ e^{z^2}
\end{equation}

$I(0)$ is obtained by calculating directly the integral:
\begin{equation}
I(0) = 2x_0 \int_{0}^{+\infty} d x \frac{1}{x^2-x_0^2} = i \pi \text{sign}(\Im z_0)
\end{equation}

And finally, anticipating that $\text{sign}(\Im z_0) = -1$ for our later case, we have
\begin{equation}
I = I(1) = -i \pi e^{-z_0^2} - \pi e^{-z_0^2} \text{erfi}(z_0)
\end{equation}

\subsection{Application to SXRS}

Returning back to $I_2(\Omega)$, we first write the product of the two Gaussian as a single one:
\begin{equation}
\frac{1}{\sqrt{2\pi}\sigma_2}e^{-(\omega_2-\omega_c)^2/2\sigma_2^2}
\frac{1}{\sqrt{2\pi}\sigma_2}e^{-(\omega_2-\omega_c-\Omega)^2/2\sigma_2^2}
=
\frac{e^{-\Omega^2/4\sigma_2^2}}{2\pi\sigma^2}
e^{-(\omega_2-(\omega_c+\Omega/2))^2/\sigma_2^2}
\end{equation}

In $I_2(\Omega)$, we make the change of variable $z = (\omega_2-(\omega_c+\Omega/2))/\sigma_2$, $d\omega_2 = \sigma_2 dz$ and defining
\begin{equation}
z_0 = -\frac{1}{\sigma_2}(\omega_c+\Omega/2-\omega_t+i \epsilon)
\end{equation}

Then, the integral can be carried out using Eq. 9, and we get:
\begin{equation}
I_2(\Omega) = - |E_2|^2 \frac{e^{-\Omega^2/4\sigma_2^2}}{2\sigma^2}e^{-z_0^2}(i+\text{erfi}(z_0))
\end{equation}

Putting this into the signal definition, we get the final signal expression:
\begin{multline}
S_\text{SXRS}(\Omega) = |E_2|^2|E_1|^2\frac{2}{\hbar^4}\Im\sum_{ecc'}
\bm \mu_{gc'}\bm \mu^\dagger_{c'e}\bm \mu_{ec}\bm \mu^\dagger_{cg}
\frac{I_{2,c'g}(\Omega)I_{1,cg}(\Omega)}{\Omega-\omega_{eg}+i\epsilon}
\end{multline}

\noindent with
\begin{multline}
I_{2,c'g}(\Omega) = \frac{e^{-\Omega^2/4\sigma_2^2}}{2\sigma_2^2}e^{-z_2^2}(i+\text{erfi}(z_0)) \ \ \ \ \ \ \ \text{with} \ \ \ \ \ \ \ z_2 = -\frac{1}{\sigma_2}(\bar\omega_2+\Omega/2-\omega_{c'g}+i \epsilon)
\end{multline}

\begin{multline}
I_{1,cg}(\Omega) = \frac{e^{-\Omega^1/4\sigma_2^2}}{2\sigma_1^2}e^{-z_1^2}(i+\text{erfi}(z_1)) \ \ \ \ \ \ \ \text{with} \ \ \ \ \ \ \ z_1 = -\frac{1}{\sigma_1}(\bar\omega_1+\Omega/2-\omega_{cg}+i \epsilon)
\end{multline}

\section{Expressions for 2DES signals}
\subsection{The impulsive limit}

The signals expressions can be further simplified by assuming impulsive pulses. Signals are then given by:
\begin{multline}
S_{\text{achir}}(T_3,T_2,T_1) = -\frac{2}{\hbar} \Im\bold R_{\mu\mu\mu\mu}(T_3,T_2,T_1) \bullet (\bm \epsilon_s\otimes \bm \epsilon_3 \otimes \bm \epsilon_2 \otimes \bm \epsilon_1)
\label{heterodynenonchir}
\end{multline}
\begin{eqnarray}
&& S_{\text{chiral}}(\Gamma) =  -\frac{2}{\hbar c}\Im \nonumber\\
&&( -\bm R_{m\mu\mu\mu} \bullet \bm b_s\otimes \bm \epsilon_3 \otimes \bm \epsilon_2 \otimes \bm \epsilon_1
- \bm R_{q\mu\mu\mu}\bullet (i \omega_s\bm{\hat k}_s \otimes \bm \epsilon_s)\otimes \bm \epsilon_3 \otimes \bm \epsilon_2 \otimes \bm \epsilon_1\nonumber\\
&&+ u_3\bm R_{\mu m\mu\mu} \bullet (\bm\epsilon_s\otimes \bm b_3 \otimes \bm\epsilon_2 \otimes \bm\epsilon_1) +u_3 \bm R_{\mu q\mu\mu}\bullet (\bm\epsilon_s\otimes (i\omega_3\bm{\hat k}_3 \otimes \bm\epsilon_3) \otimes \bm\epsilon_2 \otimes \bm\epsilon_1)\nonumber\\
&&+u_2\bm R_{\mu \mu m\mu} \bullet (\bm\epsilon_s\otimes \bm\epsilon_3 \otimes \bm b_2 \otimes \bm\epsilon_1) 
+u_2 \bm R_{\mu\mu q\mu}\bullet (\bm\epsilon_s\otimes \bm\epsilon_3 \otimes (i \omega_2\bm{\hat k}_2 \otimes \bm\epsilon_2) \otimes \bm\epsilon_1)\nonumber\\
&&+u_1\bm R_{\mu\mu\mu m} \bullet (\bm\epsilon_s\otimes \bm\epsilon_3 \otimes \bm\epsilon_2 \otimes \bm b_1) 
+ u_1\bm R_{\mu\mu\mu q}\bullet (\bm\epsilon_s\otimes \bm\epsilon_3 \otimes \bm\epsilon_2 \otimes (i\omega_1\bm{\hat k}_1 \otimes\bm\epsilon_1))
\label{chiraltot1}
\end{eqnarray}
\noindent where we have introduced the magnetic field polarization $\bm b_i = \bm{\hat k}_i \wedge \bm \epsilon_i$.


\section{Sum-Over-States expressions for 2DES}

The 2DES signal is usually split into two contributions, the rephasing ($\bm k_\text{I}$ technique) and the non-rephasing ($\bm k_\text{II}$ technique) signals.

The rephasing signal is given by
\begin{multline}
S_\text{2DES}^\text{rephasing}(T_1,T_2,T_3) = \frac{2}{\hbar^4}\Re \int dtdt_3dt_2dt_1  \\
\bm E_\text{s}^*(t)\bm E_3(t-t_3+T_3)\bm E_2(t-t_3-t_2+T_3+T_2)\bm E_1^*(t-t_3-t_2-t_1+T_3+T_2+T_1)
\\
\times\Big(
\langle\langle \bm \mu | \mathcal G(t_3)\bm \mu_\text{left}^\dagger \mathcal G(t_2)\bm \mu_\text{right}^\dagger\mathcal G(t_1)\bm \mu_\text{right}|\rho_{-\infty}\rangle\rangle
+\langle\langle \bm \mu | \mathcal G(t_3)\bm \mu_\text{right}^\dagger \mathcal G(t_2)\bm \mu_\text{left}^\dagger\mathcal G(t_1)\bm \mu_\text{right}|\rho_{-\infty}\rangle\rangle\\
-\langle\langle \bm \mu | \mathcal G(t_3)\bm \mu_\text{left}^\dagger \mathcal G(t_2)\bm \mu_\text{left}^\dagger\mathcal G(t_1)\bm \mu_\text{right}|\rho_{-\infty}\rangle\rangle
\Big)
\end{multline}

In the impulsive limit, each pulse is given by:
\begin{eqnarray}
\bm E_\text{s}^*(t) &=& \delta(t) \bm e_s^* \ e^{i \omega_s t}\\
\bm E_3(t-t_3+T_3) &=& \delta(t-t_3+T_3) \bm e_3 \ e^{-i \omega_3 (t-t_3+T_3)}\\
\bm E_2(t-t_3-t_2+T_3+T_2) &=& \delta(t-t_3-t_2+T_3+T_2) \bm e_2 \ e^{-i \omega_2 (t-t_3-t_2+T_3+T_2)}\\
\bm E_1^*(t-t_3-t_2-t_1+T_3+T_2+T_1) &=& \delta(t-t_3-t_2-t_1+T_3+T_2+T_1) \bm e_1^* \ e^{i \omega_1 (t-t_3-t_2-t_1-T_3-T_2-T_1)}
\end{eqnarray}

Summing over states and Fourier transforming over $T_1$ and $T_3$ leads to
\begin{multline}
S_\text{2DES}^\text{rephasing}(\Omega_1,T_2,\Omega_3) = -\frac{2}{\hbar^4}\Re
\bm e_s^* \bm e_3 \bm e_2\bm e_1^* \Big(
\frac{\bm \mu_{g'e'}\bm \mu_{e'g}\bm \mu_{g'e}\bm \mu_{eg} e^{- i\omega_{gg'} T_2 - \Gamma_{gg'} T_2}}
{(-\Omega_3- \omega_{e'g'} + i \Gamma_{e'g'})(-\Omega_1+\omega_{eg} + i \Gamma_{eg})}\\
+
\frac{\bm \mu_{g'e'}\bm \mu_{g'e}\bm \mu_{e'g}\bm \mu_{eg} e^{- i\omega_{e'e} T_2 - \Gamma_{e'e} T_2}}
{(-\Omega_3- \omega_{e'g'} + i \Gamma_{e'g'})(-\Omega_1+\omega_{eg} + i \Gamma_{eg})}
-
\frac{\bm \mu_{ef}\bm \mu_{fe'}\bm \mu_{e'g}\bm \mu_{eg} e^{- i\omega_{e'e} T_2 - \Gamma_{e'e} T_2}}
{(-\Omega_3- \omega_{fe} + i \Gamma_{fe})(-\Omega_1+\omega_{eg} + i \Gamma_{eg})}
 \Big)
\end{multline}


The non-rephasing signal is given by
\begin{multline}
S_\text{2DES}^\text{non-rephasing}(T_1,T_2,T_3) = \frac{2}{\hbar^4}\Re \int dtdt_3dt_2dt_1  \\
\bm E_\text{LO}^*(t)\bm E_3(t-t_3-T_3)\bm E_2(t-t_3-t_2-T_3-T_2)\bm E_1^*(t-t_3-t_2-t_1-T_3-T_2-T_1)
\\
\times\Big(
\langle\langle \bm \mu | \mathcal G(t_3)\bm \mu_\text{left}^\dagger \mathcal G(t_2)\bm \mu_\text{left}\mathcal G(t_1)\bm \mu_\text{left}^\dagger|\rho_{-\infty}\rangle\rangle
+\langle\langle \bm \mu | \mathcal G(t_3)\bm \mu_\text{right}^\dagger \mathcal G(t_2)\bm \mu_\text{right}\mathcal G(t_1)\bm \mu_\text{left}^\dagger|\rho_{-\infty}\rangle\rangle\\
-\langle\langle \bm \mu | \mathcal G(t_3)\bm \mu_\text{left}^\dagger \mathcal G(t_2)\bm \mu_\text{right}\mathcal G(t_1)\bm \mu_\text{left}^\dagger|\rho_{-\infty}\rangle\rangle
\Big)
\end{multline}

\begin{multline}
S_\text{2DES}^\text{rephasing}(T_1,T_2,T_3) = -\frac{2}{\hbar^4}\Re
\bm e_s^* \bm e_3 \bm e_2^*\bm e_1 \Big(
\frac{\bm \mu_{ge'}\bm \mu_{e'g'}\bm \mu_{g'e}\bm \mu_{eg} e^{- i\omega_{g'g} T_2 - \Gamma_{g'g} T_2}}
{(-\Omega_3- \omega_{e'g} + i \Gamma_{e'g})(-\Omega_1-\omega_{eg} + i \Gamma_{eg})}\\
+
\frac{\bm \mu_{g'e}\bm \mu_{g'e'}\bm \mu_{e'g}\bm \mu_{eg} e^{- i\omega_{ee'} T_2 - \Gamma_{ee'} T_2}}
{(-\Omega_3- \omega_{eg'} + i \Gamma_{eg'})(-\Omega_1-\omega_{eg} + i \Gamma_{eg})}
-
\frac{\bm \mu_{e'f}\bm \mu_{fe}\bm \mu_{e'g}\bm \mu_{eg} e^{- i\omega_{ee'} T_2 - \Gamma_{ee'} T_2}}
{(-\Omega_3- \omega_{fe'} + i \Gamma_{fe})(-\Omega_1-\omega_{eg} + i \Gamma_{eg})}
 \Big)
\end{multline}
